\documentclass[prl,twocolumn,showpacs,superscriptaddress]{revtex4}

\usepackage{graphicx,color,epsfig}
\usepackage{epsfig}
\usepackage{graphics}
\usepackage{dcolumn}
\usepackage{bm}
\usepackage{amssymb}
\usepackage{fancyhdr}

\begin{document}

\title{A cascade of magnetic field induced spin transitions in LaCoO$_3$}

\author{M.~M.~Altarawneh}
\affiliation{National High Magnetic Field Laboratory, Los Alamos National Laboratory, Los Alamos, NM 87545}
\affiliation{Department of Physics, Mu'tah University, Mu'tah, Karak, 61710, Jordan}
\author{G.-W.~Chern}
\affiliation{Theory Division, T-4 and CNLS, Los Alamos National Laboratory, Los Alamos, NM 87545}
\author{N.~Harrison}
\affiliation{National High Magnetic Field Laboratory, Los Alamos National Laboratory, Los Alamos, NM 87545}
\author{C.~D.~Batista}
\affiliation{Theory Division, Los Alamos National Laboratory, Los Alamos, NM 87545}
\author{A.~Uchida}
\affiliation{National High Magnetic Field Laboratory, Los Alamos National Laboratory, Los Alamos, NM 87545}
\author{M.~Jaime}
\affiliation{National High Magnetic Field Laboratory, Los Alamos National Laboratory, Los Alamos, NM 87545}
\author{D.~G.~Rickel}
\affiliation{National High Magnetic Field Laboratory, Los Alamos National Laboratory, Los Alamos, NM 87545}
\author{S.~A.~Crooker}
\affiliation{National High Magnetic Field Laboratory, Los Alamos National Laboratory, Los Alamos, NM 87545}
\author{C.~H.~Mielke}
\affiliation{National High Magnetic Field Laboratory, Los Alamos National Laboratory, Los Alamos, NM 87545}
\author{J.~B.~Betts}
\affiliation{National High Magnetic Field Laboratory, Los Alamos National Laboratory, Los Alamos, NM 87545}
\author{J.~F.~Mitchell}
\affiliation{Material Science Division, Argonne National Laboratory, Argonne IL 60439}
\author{M.~J.~R.~Hoch}
\affiliation{National High Magnetic Field Laboratory, Florida State University, Tallahassee, FL 32310}

\begin{abstract}

We present magnetization and magnetostriction studies of the insulating perovskite LaCoO$_3$ in magnetic fields approaching 100~T. In marked contrast with expectations from single-ion models, the data reveal \emph{two} distinct first-order spin transitions and well-defined magnetization plateaux. The magnetization at the higher plateau is only about half the saturation value expected for spin-1 Co$^{3+}$ ions. These findings strongly suggest collective behavior induced by strong interactions between different electronic -- and therefore spin -- configurations of Co$^{3+}$ ions. We propose a model of these interactions that predicts crystalline spin textures and a cascade of four magnetic phase transitions at high fields, of which the first two account for the experimental data.
\end{abstract}

\maketitle


Spin state crossovers induced by changes in the electronic configuration of transition metal ions can dramatically
alter several materials properties~\cite{konig1,gaspar1}. One example is the pressure induced spin transition of ferric ions in magnesium silicate perovskite that occurs in the Earth's lower mantle~\cite{antonangeli1,hsu1}.
Apart from being a major constituent of the Earth's mantle, perovskites  exhibit a variety of behaviors which include colossal magnetoresistance~\cite{dagotto1,salamon1} and high temperature superconductivity~\cite{schrieffer1}. For certain insulating perovskites containing Fe or Co ions, the two lowest energy  multiplets of the $3d$ electrons can be separated by a small energy gap, $\Delta$, owing to  competition  between Hund's coupling and crystal field splitting. A external magnetic field, $H$, can change the electronic configuration if the  lowest energy multiplet has lower spin  than  the first excited multiplet. In this situation, structural  transitions with  unusually large magneto-elastic responses can be induced for  $ H \simeq \Delta / g \mu_B$ (where $\mu_B$ is Bohr's magneton and $g$ is the g-factor).

The energy gap $\Delta$ separating low- and high-spin electronic configurations of $3d$ ions in perovskites 
is typically of order 10-1000 meV. Controlled switching between these electronic  configurations with magnetic fields therefore requires extremely large fields of order 100-10000~T (assuming $g\approx2$). Therefore, such studies have proven difficult to date. The recent development of nondestructive pulsed magnets with peak fields of 100~T and with long (ms-timescale) pulse durations now provides an opportunity to realize field-tuned spin state transitions in compounds with $\Delta \sim 12$ meV. Such studies are highly desirable, as they would provide an essential complement to pressure-induced spin cross-over phenomena that have been studied in a number of materials~\cite{konishi1,miller1,oka1}. Moreover, high-field studies allow one to test whether single-ion models suffice to describe spin-crossover phenomena in perovskites, or whether interactions \emph{between} $3d$ ions are important.

To this end we study the insulating perovskite cobaltite (LaCoO$_3$, or LCO), whose octahedrally-coordinated Co$^{3+}$ ($3d^6$) ions are natural candidates to explore field-induced transitions of the electronic configuration and spin ~\cite{raccah1, johnsson1}. A small energy gap $\Delta\approx$~12~meV separating the spin singlet ground state (6 electrons in the $t_{2g}$ orbitals) from 
the lowest energy magnetic configuration
results from the competition between Hund's coupling and crystal field splitting~\cite{abbate1}.  While the  Co$^{3+}$ ions are in their $S = 0$ state at low temperatures, thermal activation to a magnetic $S \neq 0$ state  occurs above $\sim$30 K, giving rise to a paramagnetic response. Although considerable work has been carried out on this thermally-induced spin crossover~\cite{korotin,hoch,hozoi1}, the spin value $S$ of the lowest energy excited multiplet is still controversial. A single-ion model allows to explain the temperature dependence of the magnetic susceptibility and the measured g-factor ($g =3.4$) with an $S =1$ triplet~\cite{hoch}. However, field-dependent magnetization studies have recently suggested that interactions between Co ions may also play a crucial role in LCO: a field-induced gap-closing study shows a jump in the magnetization $M$ to a plateau value of $\simeq 0.5 \mu_B$ just above 60 T~\cite{sato}. This value is much lower than the $2\mu_B$ ($4\mu_B$)  expected from saturated $S=1$ ($S=2$) Co$^{3+}$ ions within a single-ion model.  From now on we will assume that $S=1$, because this assumption leads to a simpler explanation of the observed plateaux.

Very recently Platonov {\it et al.}~\cite{Platonov12} have reported the detection of several magnetic field-induced phase transitions in LCO single crystals at 4.2K. The samples are subjected to very high magnetic fields ($\approx$~500 T) produced by rapid ($\approx$~15 $\mu$s) explosive compression of magnetic flux. The results show that the magnetization per Co$^3+$ ion starts to rise significantly for $H > 50$T, where $M/n_{Co} \simeq 0.4 \mu_B$, reaching a plateau value of $M/n_{Co} \simeq 1.4 \mu_B$  at 140 T, and a maximum value of 3.5$\mu_B$ at 500 T. It is suggested that the observed plateau and other features may be linked to antiferromagnetic interactions.\cite{Platonov12} Smoothing of the first order transition near 60 T~\cite{sato}  is attributed to the rapid variation of the magnetic field. No firm conclusions on the Co$^{3+}$ spin state are presented, although it is suggested that the $S=2$ state may be important at the highest fields reached.

Here we measure the magnetization ($M$) of LCO in longer-duration non-destructive pulsed magnetic fields approaching 100 T \cite{feder1}. In contrast to expectations from single-ion models, we observe multiple magnetization steps and plateaux, which indicate the relevance of inter-ion coupling. Moreover, magnetostriction studies complement the magnetization results, and reveal large lattice changes induced by a combination of two factors: a) the $S$=0 and $S$=1 electronic configurations of the Co$^{3+}$ ions have different volume and b) the $S$=1 configuration is Jahn-Teller active. Susceptibility measurements made at higher fields in a single turn pulsed field experiment reveal that the second plateau persists at least up to 140~T. We propose a  model that allows for collective behavior of the Co$^{3+}$ ions and reveals a rich phase diagram in which the  plateaux originate from different stable crystalline textures of $S=1$ Co$^{3+}$ ions in a background of $S=0$ configurations. The orbital degree of freedom of the $S=1$ Co ions also leads to orbital ordering that should coexist with the magnetic ordering.
\begin{figure}
\includegraphics[width=0.95\columnwidth]{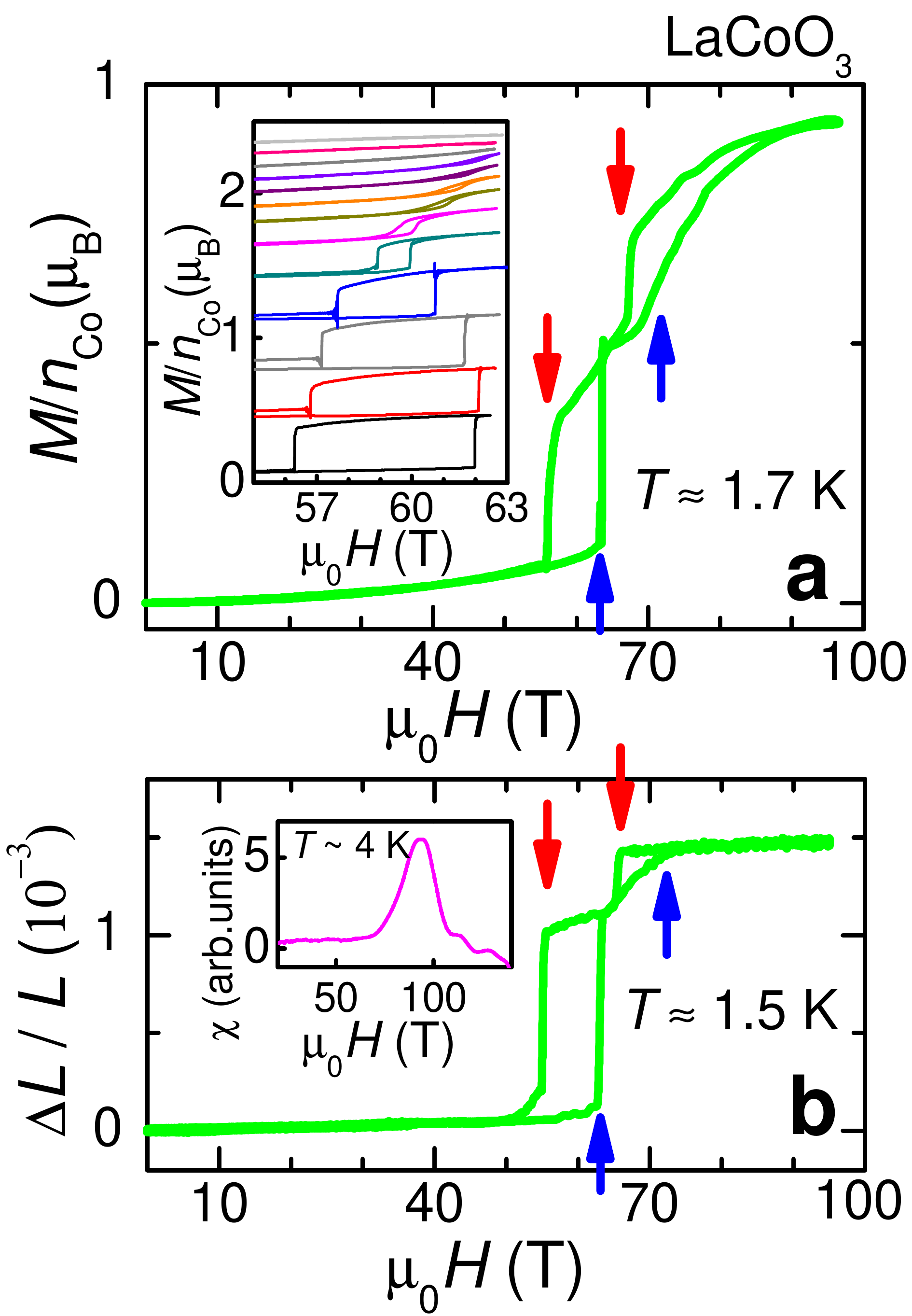}
\caption{(Color online) (a) Magnetization of LaCoO$_3$ in units $\mu_B$/Co as a function of applied field at $T =1.7$ K.
The blue and red arrows indicate transitions for increasing and decreasing fields respectively. Two plateau regions can be seen.
The inset shows an expanded view of the first transition between 55 and 63 T for $T = 1.6$ (bottom), 4.0, 10.0, 20.0, 25.0, 27.0,
28.0, 28.5, 29.0, 30.0, 40 and 55 (top) K. (b) Magnetostriction of LCO, measured using a high sensitivity optical strain gauge,
as a function of applied field. The inset shows the susceptibility $\chi = dM/dH$ measured in a pulsed single
turn coil (140 T in $\sim 2$ $\mu$s). The two transitions are merged into a single peak in the very rapidly
changing field. No further transitions are found below 140 T.}
\label{data}
\end{figure}

The single crystals of LCO are grown in a floating zone furnace at Argonne National Laboratory, with
pieces being cut and polished into either small tapered cylinders for the magnetization $M$ measurements or $1\times1\times3$mm bar shaped samples
for the magnetostriction experiments. Magnetization curves are obtained as a function of $\mu_0 H$, using an extraction
magnetometer~\cite{detwiler1}, with pulsed fields up to 97 T at temperatures in the range  1.6--55 K. High precision magnetostriction measurements are made using a fiber-optic strain gauge~\cite{daou1,sup}. In all experiments, singe crystals are oriented with their [010] pseudo-cubic axes parallel to $H$.

Figure~\ref{data}(a) shows $M$ versus $\mu_0 H$ at $T =1.7$ K revealing two transitions, each accompanied by
increases $\Delta M \sim 0.5\,\mu_B$/Co, and two plateau regions consistent with two distinct magnetic phases. The second
plateau occurs at $M \sim 0.9\,\mu_B$/Co which is well below the magnetization of 2 $\mu_B$/Co expected from saturated Co ions. Both transitions exhibit
hysteresis in this pulsed field experiment. The location of the transition between rising and falling fields, as indicated by the
blue and red arrows, points to a diffusionless, displacive transition type~\cite{khachaturyan,cowley,buerger1,hyde1}.
The inset in Fig.~\ref{data}(a) for fields in the range 55--63 T, and with offsets for clarity, presents an expanded view of
the first transition at temperatures between 1.6 K and 55 K. The hysteresis diminishes as $T$ is raised. Once initiated,
the transitions are observed to proceed rapidly, within $\sim 5$ $\mu$s, over the entire sample (length $\sim$ 4 mm).
Figure~\ref{data}(b) shows the magnetostriction measured parallel to the field. The transition fields and $\Delta L/L$ plateaux closely match the magnetization behavior in Fig.~\ref{data}(a).
The red and blue arrows again represent rising and falling field transitions. The magnitude of the field-induced
strain $\Delta L/L \sim 10^{-3}$ exceeds that found in structurally related perovskites~\cite{asamitsu1}. For the rapidly rising fields, some rounding of the transitions is attributed
to sample heating associated with latent heat released at the transitions, which occurs despite immersion in superfluid $^4$He.
The magnetostriction results, which show sample expansion $(\partial V/\partial H)_{T,P} > 0$ ,
are consistent with the Maxwell relations since $(\partial M/\partial P)_{T,H} < 0$ for $T < 100$ K~\cite{daou1}.
The inset in Fig.~\ref{data}(b) shows the magnetic susceptibility at 4 K measured in a pulsed single turn magnet
which reaches $\sim 140$ T.

\begin{figure}[t]
\includegraphics[width=0.95\columnwidth]{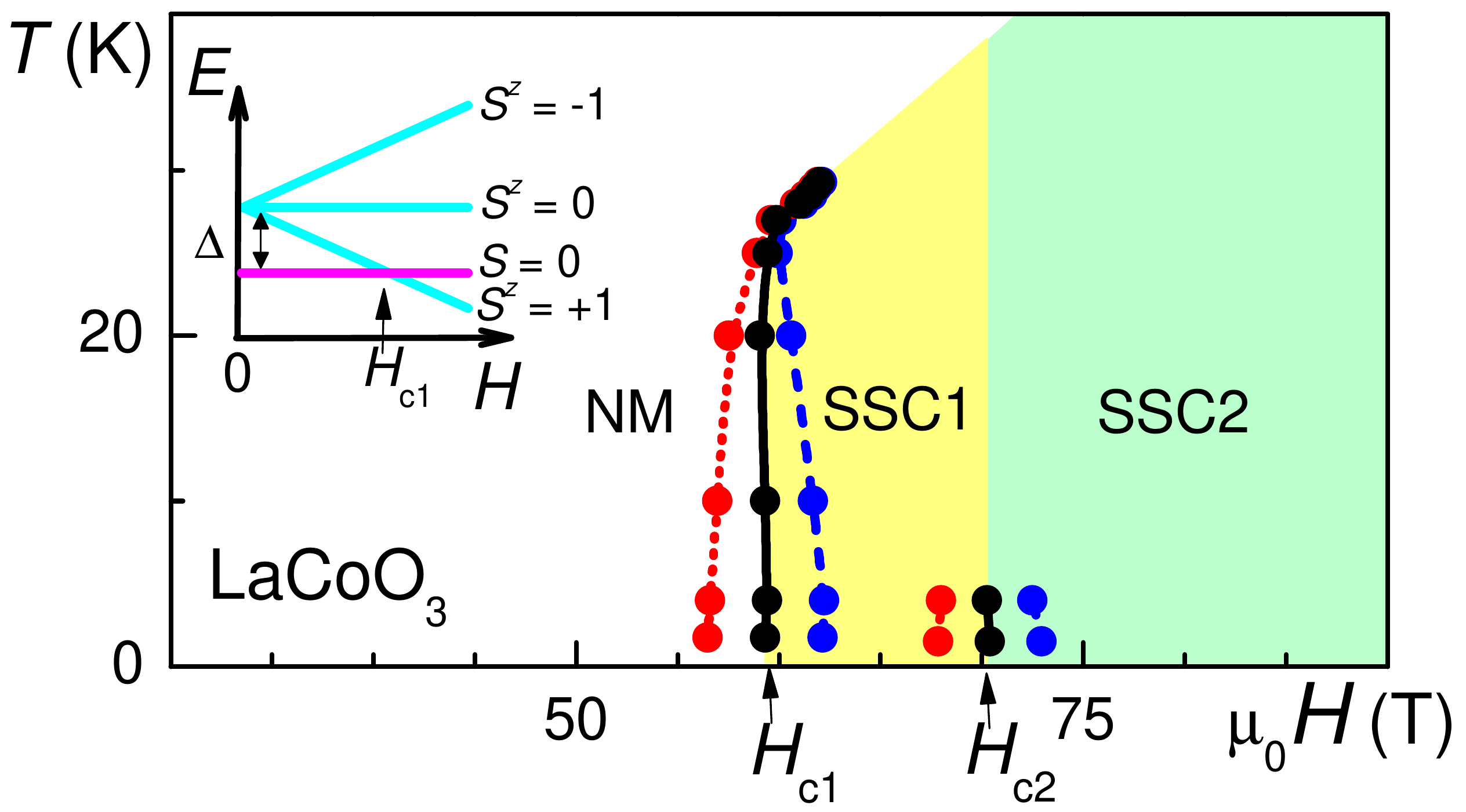}
\caption{(Color online) Phase diagram for LCO based on the results shown in Fig.~\ref{data}(a). Blue and red symbols denote rising
and falling fields respectively, while black symbols connected by the solid curve represent the thermodynamic phase boundary
obtained as an average of the rising and falling field values at a given $T$. SSC1 and SSC2 denote spin state crystalline one and two respectively. The inset shows low lying energy states as a
function of applied field for Co$^{3+}$ in LCO based on a single ion description~\cite{sato}.}
\label{phasediagram}
\end{figure}

Figure~\ref{phasediagram} shows a phase diagram for LCO based on the inset curves in Fig.~\ref{data}.
Besides the non-magnetic (NM) phase at low fields, there are two ordered magnetic phases denoted spin state crystalline 1 (SSC1)
and 2 (SSC2) respectively. The inset depicts low-lying single-ion states for Co$^{3+}$ in LCO~\cite{hoch,sato}.

The observed field induced phase transitions must involve interactions between the non-magnetic ions and the magnetic $S^z=$~+1 ionic configurations that 
are stabilized by the field. We note that the magnetic ions  are both larger in volume as well as Jahn-Teller active~\cite{hoch,sato}. Indeed, the $S=+1$ Co$^{3+}$ ion with a $t_{2g}^5 e_g^1$ electron configuration has an orbital degree of freedom since there is one electron shared between two $e_g$ orbitals. We assume that the energy gap ($\Delta$) between the $S=1$ and nonmagnetic configurations is large in comparison to the exchange interactions between different magnetic configurations of Co$^{3+}$ ions. We also assume that the gap to the $S = 2$ magnetic configuration is much larger than $2\Delta$ and consequently this configuration can be eliminated from the low-energy Hilbert space for fields up to 100 T. This assumption is supported by the consistency of the model predictions with experiment. Applying a magnetic field $H$ lowers the energy of the doubly degenerate $S^z=+1$ states (the $e_g$ orbitals being $d_{3z^2-r^2}$ and $d_{x^2-y^2}$) such that their energy becomes comparable to the $S=0$ nonmagnetic state. A low-energy effective model is obtained by projecting the original Hamiltonian into the subspace of these three lowest-energy states.

The orbital state of the $S^z=+1$ ion is parameterized by an angle $\theta$: $|\theta\rangle=\cos(\theta/2)|3z^2-r^2\rangle+\sin(\theta/2)|x^2-y^2\rangle$, corresponding to a biaxial deformation of the local O$_6$ octahedra. The doublet orbital degeneracy manifested as a continuous symmetry for the angle $\theta$ is retained by linear Jahn-Teller (JT) interaction. In real compounds, however, the continuous degeneracy
is lifted by the lattice anharmonicity and higher-order interactions~\cite{kk03}. In fact, only local elongations of the O$_6$ octahedra are observed in almost all JT ions with $e_g$ electrons~\cite{kk82}. Depending on the local elongation axis, the allowed orbitals are $d_{3x^2-r^2}$, $d_{3y^2-r^2}$ and $d_{3z^2-r^2}$, corresponding to $\theta = 0$ and $\pm 2\pi/3$. The large nonlinear JT distortion also lifts the degeneracy of the $t_{2g}$ levels and quenches the
orbital degrees of freedom associated with the hole left in the $t_{2g}$ manifold of the $S^z=1$ state.

A single $S=1$ ion embedded in the host matrix acquires additional energy via JT coupling with the local octahedron and exchange interaction with its non-magnetic neighbors, amounting to a slight modification ($\tilde{\Delta}$) of the energy gap:
\begin{eqnarray}\label{singleion}
	\mathcal{H}_1 = \sum_i \left(\tilde\Delta - g\mu_B H\right) n_i.
\end{eqnarray}
Here $n_i=S^z_i=+1$ such that $n_i=$~1 for a magnetic configuration on the ion $i$, $\tilde\Delta$ is the renormalized spin state gap and $g\mu_B H$ is the Zeeman interaction. The single-ion physics described by Eq.~(\ref{singleion}) implies a single field-induced crossover when $g\mu_B H > \tilde \Delta$. To account for the observed multiple transitions, we need to include the interactions between the $S^z=+1$ sites:
\begin{eqnarray}
	\label{eq:H2}
	\mathcal{H}_2 = \frac{1}{2}\,\sum_{i,j}\, \Bigl[ V_{ij}
	+ V'_{ij}(s_i, s_j) \Bigr] n_i n_j.
\end{eqnarray}
The leading order coupling ($V_{ij}$) is isotropic and repulsive because it results from the increased volume of the $S^z=1$ ions relative to the nonmagnetic ones. The second interaction terms are orbital-dependent and determine the relative orbital orientations of the magnetic ions.~\cite{sup} The  three-state Potts variable $s_i$  a  indicates the orbital states ($d_{3l^2-r^2}$ with $l=x$, $y$, or $z$) of the $S^z=1$ ion.

The origin of the isotropic interaction $V_{ij}$ can be understood by using the so called ``sphere-in-the-hole'' model~\cite{kk03}. The larger ionic size of a $S=1$ ion in the host matrix of $S=0$ ions acts as an elastic impurity and creates a strain field which decays  as $\sim 1/r^3$ at large distances. A second  impurity interacts with this strain field, giving rise to the elastic term in Equation~(\ref{eq:H2}).

\begin{figure}[t]
\includegraphics[width=0.95\columnwidth]{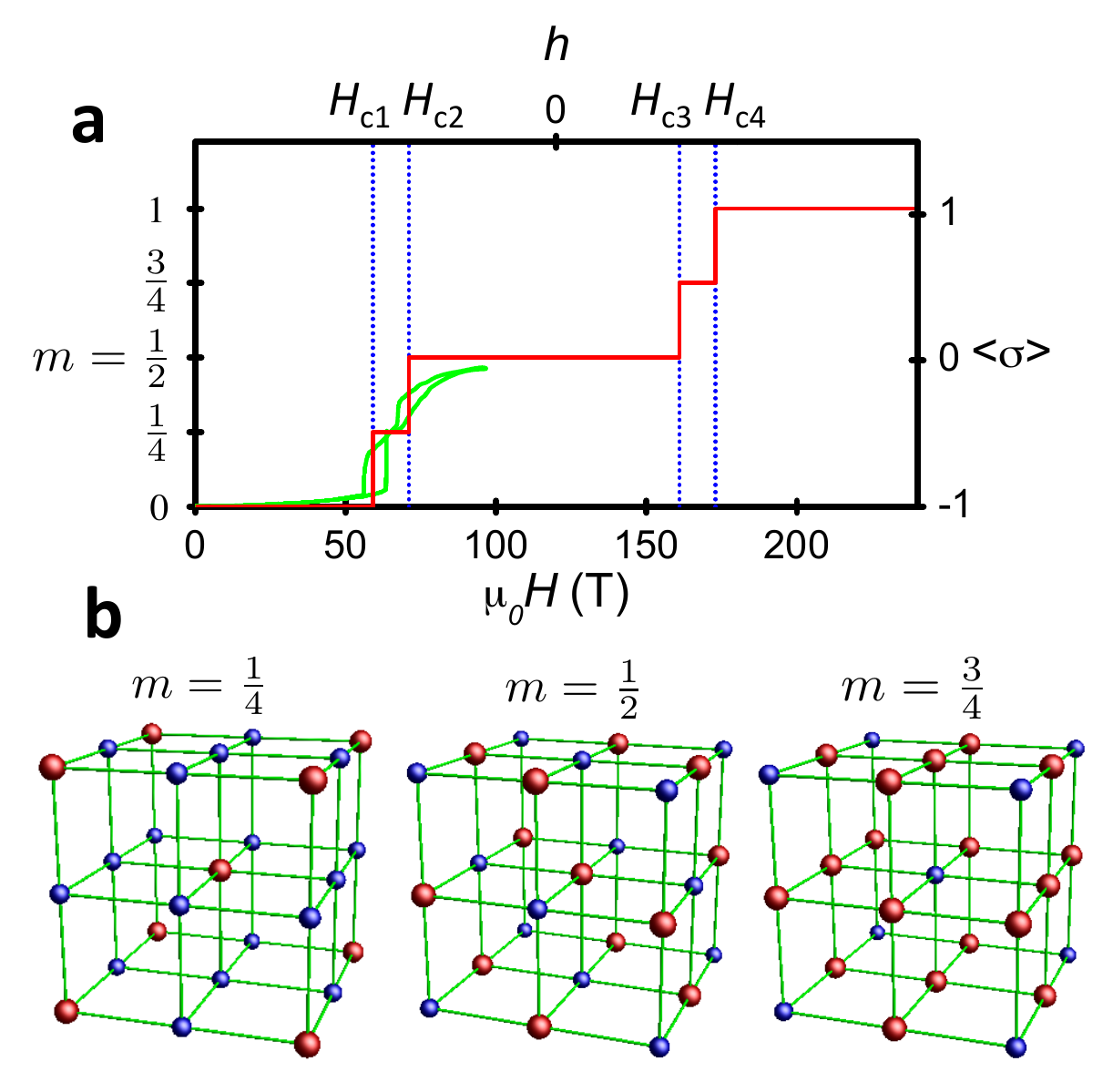}
\caption{(Color online) (a) The red curve gives the predicted normalized magnetization $m$ versus $\mu_0 H$ behavior for LCO based
on the interacting spin model. A cascade of four transitions is shown at fields $H_{c1}$ to $H_{c4}$ as described in the text.
The scaled green curve is based on the experimental results shown in Fig.~\ref{data}(a). The right hand axis gives values of
the pseudospin variable $\langle \sigma \rangle$ used in the theoretical model. (b) Predicted spin state crystalline structures
for each of the phases with fractional magnetization plateaux in Panel~(a). The blue and (larger) red balls denote Co$^{3+}$
ions in the $S = 0$ and $S_z = +1$ states respectively. The intermediate spin Co ions form body-centered (BCC) and a face-centered
(FCC) cubic structures in the $m =1/4$ and $m = 1/2$ plateau regions respectively. Orbital ordering may induce a further
lowering of the symmetry \cite{sup}.}
\label{m-h}
\end{figure}

We find that the key aspects of the data $-$ namely multiple phase transitions, incremental steps in the magnetization, lattice expansions and metastability $-$ can be captured by considering field-induced $S^z=+1$ spin states in which we neglect orbital orientation-dependent terms. The Hamiltonian can then be mapped onto an effective Ising model
\begin{eqnarray}
	\mathcal{H}_{\rm eff} = J_1 \sum_{\langle ij \rangle} \sigma_i \sigma_j
	+J_2 \sum_{\langle \langle ij \rangle \rangle} \sigma_i \sigma_j
	- h \sum_i \sigma_i,
	\label{eq:ising}
\end{eqnarray}
on a cubic lattice with nearest neighbor and next nearest neighbor antiferromagnetic interactions $J_1=V_1/4$ and $J_2=V_2/4$, respectively, between $\sigma_i=\pm$~1 pseudospins. $S=$~0 becomes $\sigma_i=-$~1 and $S^z=$~1 becomes $\sigma_i=+$~1 via the transformation $n_i=(1+\sigma_i)/2$ and the adoption of an effective magnetic field $h=\frac{1}{2}g\mu_B H-\frac{1}{2}\tilde \Delta-\frac{3}{2}V_1-3V_2$ (shown schematically in Fig.~\ref{m-h}). We note the orbital orientation-dependent terms neglected in
Eq.(\ref{eq:ising}) are  relevant for deciding the orbital ordering that accompanies each spin ordering at low enough temperatures.

Provided the nearest neighbor coupling dominates (i.e. $J_1>4J_2$), the  $h=$~0 ground state $h=$~0 is a two-sublattice N\'eel order, whose actual magnetization is half of the fully saturated value $m=\frac{1}{2}$ corresponding to $\langle\sigma_i\rangle=$~0. We associate this ordering with the measured magnetization plateau in the region above $\sim$~75~T shown in Figs.~\ref{phasediagram} and~\ref{data}. Intermediate plateaux with $m=\frac{1}{4}$ and $m=\frac{3}{4}$ (in which $\langle\sigma_i\rangle=\mp\frac{1}{2}$) become viable for finite $J_2$. On transforming back to the original Hamiltonian (considering both positive and negative $h$), we therefore realize a cascade of first order transitions between different phases with critical fields $-$ $H_{\rm c1}$, $H_{\rm c2}$, $H_{\rm c3}$ and $H_{\rm c4}$ in Fig.~\ref{m-h} $-$ whose relative separations depend on the nearest and next nearest neighbor repulsions $V_1$ and $V_2$.

Fig.~\ref{m-h} shows quantitative consistency between the measured magnetization steps and those of the model  owing to the quenching of the orbital contribution to the $g$-factor (such that $g\sim2$) by a JT distortion of the O$_6$ octahedra encasing the $S^z=+1$ sites. On associating the first two transitions ($H_{\rm c1}$ and $H_{\rm c2}$) with the two transitions observed experimentally, we estimate $V_2\approx$~1.2~K. The lack of further transitions in motor-generator-driven magnetic fields extending to 97~T (in Fig.~\ref{data}), and  in a single turn magnet system delivering fields to $\approx$~140~T (shown in the inset to Fig.~\ref{data}(b)), implies that $V_1\gtrsim$~25~K. On enumerating $V_1$ over the six nearest neighbor Co atoms in the cubic perovskite, we arrive at a lower limit $\lesssim$~150~K for the  energy scale over which collective effects influence the phase diagram of LCO.

Since the $S^z=+1$ ions form a bipartite BCC structure, the dominant antiferro-orbital interaction gives rise to a staggered orbital order in the $m=1/4$ plateau. On the other hand, antiferro-orbital interactions are frustrated in the second $m=1/2$ plateau
as the $S^z=+1$ ions form a non-bipartite FCC lattice. Nonetheless, our Monte Carlo  simulations found a partial orbital ordering with a layered structure.

The different crystalline structures in Fig.~\ref{m-h}(b) (e.g. BCC and FCC) anticipated for each of the phases of LaCoO$_3$ provide an explanation for the discontinuous changes in the lattice and hysteretic behavior. Each  magnetization plateau in Fig.~\ref{m-h}(a) corresponds to a different optimal sublattice arrangement of $S=0$ and $S^z=+1$ spin states. Orbitals in the $S^z=+1$ sublattice are more voluminous and directional in nature, causing the lattice to expand in response to the pressure exerted by their increased density (i.e. $V_1$ and $V_2$). The reduced expansion at the second transition ($H_{\rm c2}$) suggests the predominantly repulsive interaction is at least partially compensated by an attractive antiferro-orbital interaction, which is able to afford a more efficient arrangement of the orbitals at higher densities \cite{sup}.

Finally, we reiterate that  the spin value $S$ of the first excited multiplet is still matter of debate \cite{hoch,podlesnyak06,Kyomen05}.  
While the generic form of the effective Ising Hamiltonian of Eq.~(\ref{eq:ising}) does not depend on the value of $S$, one would have to include further neighbor repulsions among magnetic configurations in order to explain a $\sim 0.5\mu_B$ magnetization plateau for $S=2$ ($\sim 0.5\mu_B$ is 1/8 of the saturated value in this case). Therefore, our original assumption of $S=1$ leads to a simpler and more natural explanation of the measured plateaux.
We also note that the orbital physics and magnetostrictive properties are likely  to be very different for $S=2$ owing to the $e_g$ orbitals no longer being JT active.

The experimental evidence therefore suggests that collective behavior involving  two spin states leads to different crystalline arrangements that can be tuned with magnetic field. While magnetostriction associated with field-tuned orbital order has been reported in  manganites~\cite{asamitsu1}, the present behavior in LaCoO$_3$ is different in that a strong coupling of spin, orbital and lattice degrees of freedom occurs in a Mott insulator. An entirely different type of functionality results: multiple field-tuned (diffusionless) transitions giving rise to a rapidly switchable strain. Spin state crystallization is likely to be a general property of crystalline materials subject to a spin state transition under extreme conditions~\cite{antonangeli1,hsu1,konishi1,miller1}, causing such materials to potentially become metastable in the vicinity of a phase transition and vulnerable to a sudden release of mechanical energy.

{\it Acknowledgements.}
Work at the National High Magnetic Field Laboratory is supported by the National Science Foundation under 
DMR-0654118, Florida State University, the State of Florida and the U.S. DOE BES project ``Science at
100 tesla". Work at Argonne National Laboratory is supported by the
U.S. Department of Energy under Contract No. DEAC02-06CH211357.

\section{supplementary information}
\subsection{Further Experimental details}
\subsubsection{Magnetization}
Magnetization to $\approx$~63~T is measured in a `short-pulse' capacitor-driven pulsed magnet over temperatures ranging between 1.7 and 55~K using an extraction magnetometer~\cite{detwiler1} consisting of concentrically arranged counterwound coils of $\approx$~1000 turns. Magnetization studies to $\approx$~97~T are performed in a generator-driven nondestructive magnet at Los Alamos National Laboratory~\cite{feder1}. The magnetic susceptibility is measured inductively as an induced voltage, with the discontinuous nature of the transitions typically yielding a signal in the form of a delta function of height $\sim$~10~V at each of the transitions. Susceptometer measurements in the 140 T single-turn magnet system~\cite{mielke1} were made using a compensated coil consisting of a few turns.

\subsubsection{Magnetostriction}
Magnetostriction measurements are made using a fiber-optic Bragg grating (FBG) attached to the sample using cyanoacrylate bond~\cite{daou1p}. The single-mode fiber contains a 1~mm long Bragg grating and is manufactured by SmartFibre, UK. The reflection of light at the Bragg wavelength shifts when the sample length changes. Optical spectra are collected with a fast 1024 pixel InGaAs infrared line array camera, which is read out at 47 kHz. Mounted on a fiber-coupled research spectrometer of focal length 500~mm with a 600~groove/mm$^{-1}$ diffraction grating, the 25~$\mu$m pixel pitch results in a dispersion, $\delta\lambda$ of 0.056~nm per pixel at 1550~nm. To illuminate the FBG, we use a superluminescent diode broadband source with an output power of 30~mW in the wavelength range 1525-1565~nm.

\subsubsection{Dissipation effects}
The hysteresis at each of the transitions evidences dissipation resulting in the unavoidable generation of heat during the experiments. This irreversible heating is likely partially compensated by cooling due to the magnetocaloric effect, which in turn is caused by the field-induced closing of the spin gap. The sample is partially able to impart the resultant heat to the surrounding cryogen medium in times of order 1~ms at base temperature ($T\approx$~1.6~K) where the sample is immersed in superfluid $^4$He. Owing to the rapid succession of the two transitions on the rising magnetic field where the rate of change of magnetic field is highest, not all of the heat generated by the first transition is able to escape the sample before the second transition, possibly contributing to the temperature-rounding of the second transition on the rising magnetic field. The slower ramp rate of the falling magnetic field enables more heat to dissipate to the cryogen between the transitions, enabling both transitions to be clearly observed with reduced rounding.

\subsubsection{Maxwell relation and field-induced volume expansion}
The Maxwell relation $\partial V/\partial H|_{T,P} = -\partial M/\partial P|_{T,H}$, where $V$ is the sample volume, $H$ is the applied magnetic field, $M$ is the sample magnetization, and $P$ is the applied hydrostatic pressure, provides a means to check our data for self consistency. We know from prior results~\cite{kozlenko} that $\partial M/\partial P <0$ for temperatures $T<100K$, caused by the increase of the crystal field gap with pressure. Hence, an increase in the sample volume  with magnetic field is expected. This is consistent with our  experimental observation of the longitudinal magnetostriction in high fields, and what we expect from our model.

\subsection{Details of the Model}

\subsubsection{Background}
In the model, we assume the single-ion ground state of Co$^{3+}$ ion to be non-magnetic ($S=0$) while the first excited state is magnetic with $S=1$. The latter has an orbital degree of freedom since there is one electron shared between two $e_g$ orbitals. We further assume that the energy gap ($\Delta$) between the $S=1$ and nonmagnetic configurations is large in comparison to the exchange interactions between different magnetic configurations of Co$^{3+}$ ions. Applying a magnetic field $H$ lowers the energy of the doubly degenerate $S^z=1$ triplet states (the $e_g$ orbitals being $d_{3z^2-r^2}$ and $d_{x^2-y^2}$) such that their energy becomes comparable to the $S=0$ nonmagnetic state. A low-energy effective model is obtained by projecting the original Hamiltonian into the subspace of the three low-energy states.

\begin{figure}
\vspace*{0cm}
\hspace*{0cm}
\includegraphics[width=0.99\columnwidth]{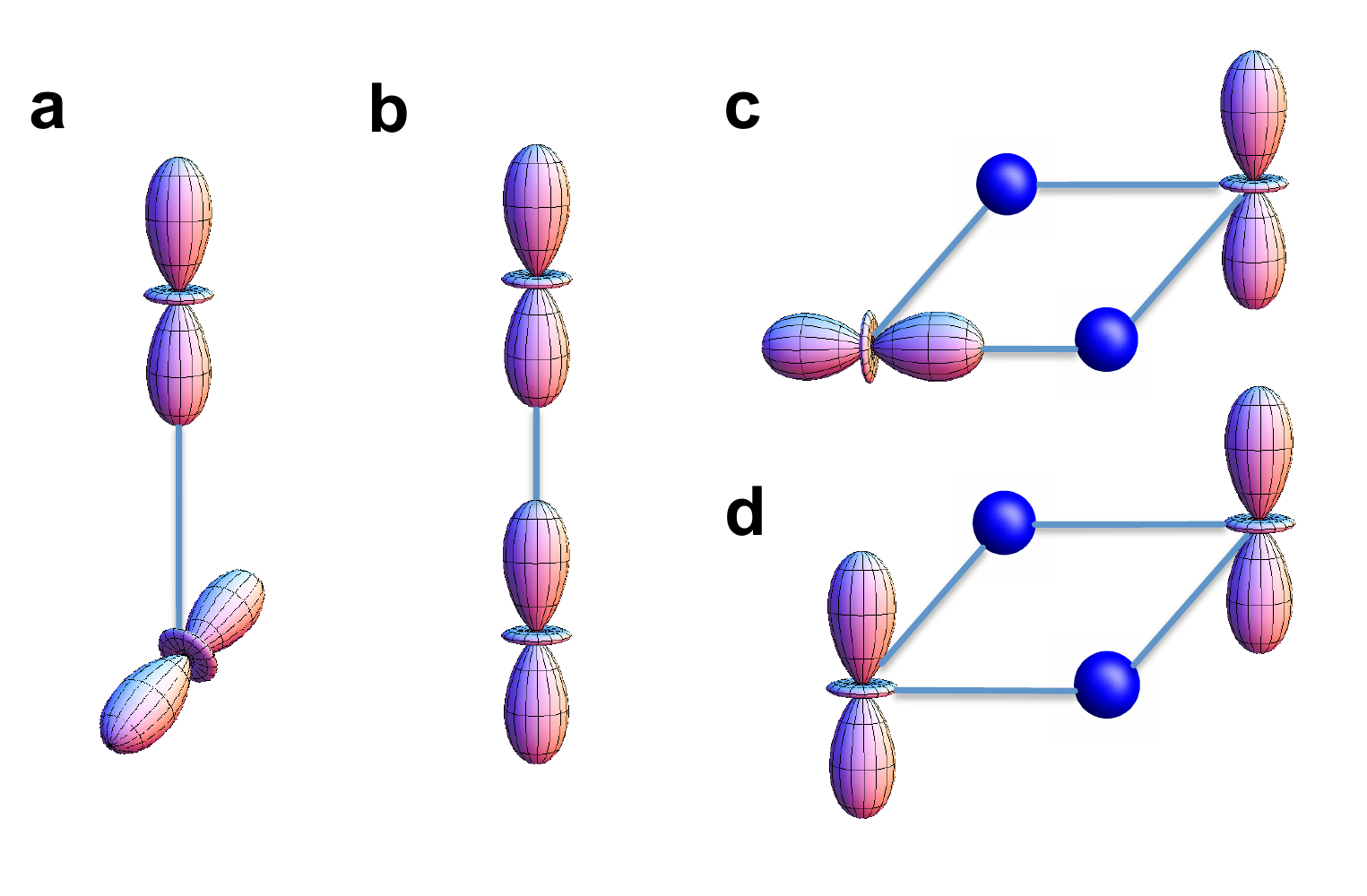}
\caption{{\bf Supplementary Information: A schematic representation of higher order interaction terms}. Antiferro and ferro-orbital configurations on a nearest-neighbor bond along $z$-axis ({\bf a} and {\bf b}) and a next-nearest-neighbor bond in the $xy$ plane ({\bf c} and {\bf d}). The corresponding interaction energies are $V^a_1$, $V^f_1$, $V^a_2$, and $V^f_2$, respectively, where the subscript `1' refers to nearest neighbor and `2' refers to second nearest neighbor.}
\label{bonds}
\end{figure}

\subsubsection{Relevant terms in the Hamiltonian}
A single $S=1$ ion embedded in the host matrix acquires additional energy via JT coupling with the local octahedron and exchange interaction with its non-magnetic neighbors, amounting to a slight modification ($\tilde{\Delta}$) of the energy gap ($\Delta$) entering into the single ion component ($\mathcal{H}_1$) of the Hamiltonian given by Eq.~(1) in the main text. As the density $\langle n_i\rangle$ of $S=1$ states increases, the interactions between them leads to the formation of spin-state superstructures and orbital order, necessitating a second component of the Hamiltonian 
\begin{eqnarray}
	\label{eq:H2}
	\mathcal{H}_2 = \frac{1}{2}\,\sum_{i,j}\, \Bigl[ V_{ij}
	+ V'_{ij}(s_i, s_j) \Bigr] n_i n_j.
\end{eqnarray}
describing interactions between different sites. $\mathcal{H}_2$ is the same as Eq.~(2) in the main text. The dominant mechanism is the elastic interaction mediated by long-wavelength phonons. Short-range interactions are usually orbital dependent and have contributions from both superexchange and JT couplings.


The origin of the isotropic ($V_{ij}$) term is discussed in the main text.
The orbital-dependent interaction potential in Eq.~(\ref{eq:H2}) consists of two terms:
\begin{eqnarray}
	V'_{ij}(s_i, s_j) = V^f_{ij}\,\mathcal{O}^f_{s_i, s_j} + V^a_{ij}\,\mathcal{O}^a_{s_i, s_j},
\end{eqnarray}
where the ferro- and antiferro-orbital bond operators are given by
\begin{eqnarray}
	\mathcal{O}^f_{s_i, s_j} &=& \delta_{s_i, \alpha_{ij}} \delta_{s_j, \alpha_{ij}}, \\
	\mathcal{O}^a_{s_i, s_j} &=& \delta_{s_i,\alpha_{ij}}\left(1-\delta_{s_j,\alpha_{ij}}\right) + 
	\left(1-\delta_{s_i,\alpha_{ij}}\right)\delta_{s_j,\alpha_{ij}}. \nonumber
	\label{eq:bond}
\end{eqnarray}
Since the orbital interaction is generally anisotropic, the variable $\alpha_{ij}=x, y, z$ indicates the orbital flavor (orientation) which has distinct symmetry properties with respect to the other two on a given bond $(ij)$. For example, $\alpha_{ij} = z$ ($d_{3z^2 - r^2}$ orbital) on a nearest-neighbor bond along $z$-axis or a next-nearest-neighbor bond in the $xy$ plane. Orbital configurations in which the $\alpha$-orbital is empty on both sides are neglected in Eq.~(\ref{eq:bond}) since such terms are irrelevant for orbital ordering.

The orbital-dependent interaction potentials ($V^f_{ij}$ and $V^a_{ij}$) result mostly from the anisotropic superexchange and JT interactions, as described in more detail in the following section. Since the elongation of O$_6$ octahedra is the source of local anisotropic stress, the resultant phonon-mediated interaction is also orbital-dependent~\cite{eremin86} thereby contributing to both $V^f_{ij}$ and $V^a_{ij}$. The interaction potentials for two $S^z=+1$ ions on the nearest neighbor and next-nearest neighbor bonds are summarized in Fig.~\ref{bonds}. The electronic contribution to the interaction potentials becomes negligible beyond a few nearest neighbors as higher-order hoppings are required for such virtual exchange processes.
The orbital-dependent interaction potentials ($V^f_{ij}$ and $V^a_{ij}$) result mostly from the anisotropic superexchange and JT interactions, as described in more detail in the following section. Since the elongation of O$_6$ octahedra is the source of local anisotropic stress, the resultant phonon-mediated interaction is also orbital-dependent~\cite{eremin86} thereby contributing to both $V^f_{ij}$ and $V^a_{ij}$. The interaction potentials for two $S^z=+1$ ions on the nearest neighbor and next-nearest neighbor bonds are summarized in Fig.~\ref{bonds}. The electronic contribution to the interaction potentials becomes negligible beyond a few nearest neighbors as higher-order hoppings are required for such virtual exchange processes.

\subsubsection{The effective Hamiltonian}
While nearest-neighbor ferro- and antiferro-orbital interactions are expected to be significant, these terms do not play a role in the formation of the first two field-induced phases. The density of $S^z=+1$ orbital states is too low to support the existence of such bonds. It is therefore reasonable to assume that the net interaction is repulsive, with larger orbitals tending to repel each other, enabling us to construct an effective Ising model Hamiltonian ($\mathcal{H}_{\rm eff}$) consisting of nearest neighbor ($V_1$) and next-nearest neighbor ($V_2$) repulsive terms as given by Eq.~(3) in the main text.

The magnetization curve $m(H)$ of the original model is shown in Fig.~3a in the main text. The width of the $m=1/4$ and 3/4 plateaus is $\Delta H_{1/4} = \Delta H_{3/4} = 24J_2 = 6 V_2$,
indicating that both states are stabilized by second-nearest neighbor interactions.

\subsubsection{Orbital ordering}
In the $m=1/4$ plateau phase, likely corresponding to the first field-induced phase (SSC1) in Fig.~3b in the main text, the closest distance between $S^z=+1$ ions is $r = \sqrt{3}a$ (i.e. third-nearest neighbors). Since superexchange and JT interactions are negligible at this distance, we resort to the anisotropic elastic interaction discussed in Ref.~\onlinecite{eremin86} to calculate the effective interaction potentials:
\begin{eqnarray}
	V^f(r) &=& 3\sigma_0\Gamma/8\pi r^3, \nonumber \\
	V^a(r) &=& -3\sigma_0\Gamma/16\pi r^3,
	\label{eq:vr}
\end{eqnarray}
where $\Gamma = (3c_{11} + 4 c_{44})/(c_{11} + 2 c_{44})$, $c_{11}$ and $c_{44}$ are elastic moduli of the crystal, and $\sigma_0$ is the strength of the local stress. Since the third-neighbor bonds are parallel to the $\langle111\rangle$ directions, the three $3l^2-r^2$ orbitals ($l=x$, $y$, $z$) are equivalent with respect to the bond. The corresponding bond operators are simply $\mathcal{O}^f_{s_i,s_j}=\delta_{s_i,s_j}$ and $\mathcal{O}^a_{s_i,s_j}=1-\delta_{s_i,s_j}$. The elastic interaction given by Equation~(\ref{eq:vr}) shows an additional energy cost for ferro-orbital configurations, we therefore expect the $S^z=+1$ ions at the $m=1/4$ plateau to form a bipartite body-centered cubic (BCC) lattice with a doubled lattice constant and a staggered  orbital order.  Since there are three possible orbital orientations, there are six different staggered orbital configurations that have the same energy in absence of spin-orbit coupling: $d_{3{\nu}^2-r^2}$, $d_{3{\nu'}^2-r^2}$ with $\nu,\nu'=\{x,y,z\}$ and $\nu\neq\nu'$.

In the $m=1/2$ plateau phase associated with the second field-induced phase (SSC2) in Fig.~3b in the main text, the next-nearest neighbor antiferro-orbital component ($V^a_2$ in Fig.~\ref{bonds}) is expected to be relevant, which has a contribution from superexchange. While indirect hopping via O ions only contributes to sixth order to the exchange energy between next-nearest-neighbor Co$^{3+}$ ions, direct hopping between $t_{2g}$ orbitals contributes to second-order. On taking into account the splitting $\delta_{\rm JT}$ of $t_{2g}$ levels due to the elongation of the local octahedron, we find
\begin{eqnarray}
	V^a_2 = - \frac{\delta_{\rm JT}\, t_{dd\sigma}^2}{U_{dd} (U_{dd} - \delta_{\rm JT})},
\end{eqnarray}
where $t_{dd\sigma}$ is the direct $dd\sigma$ hopping between $t_{2g}$ orbitals and $U_{dd}$ is the on-site Coulomb repulsion of $d$ electrons. Unlike the case of $m=1/4$ plateau, The antiferro-orbital interaction $V^a_2$ remains frustrated as the $S^z=1$ ions in the half-plateau form a non-bipartite face-centered cubic (FCC) lattice. Nonetheless, our Monte Carlo simulations found a partial orbital ordering with a layered structure. Detailed studies of the orbital order will be presented elsewhere.


\subsection{Calculation of bond energy}
\subsubsection{Exchange interaction}
\begin{figure*}
\vspace*{0cm}
\hspace*{0cm}
\includegraphics[angle=0,width=13.cm]{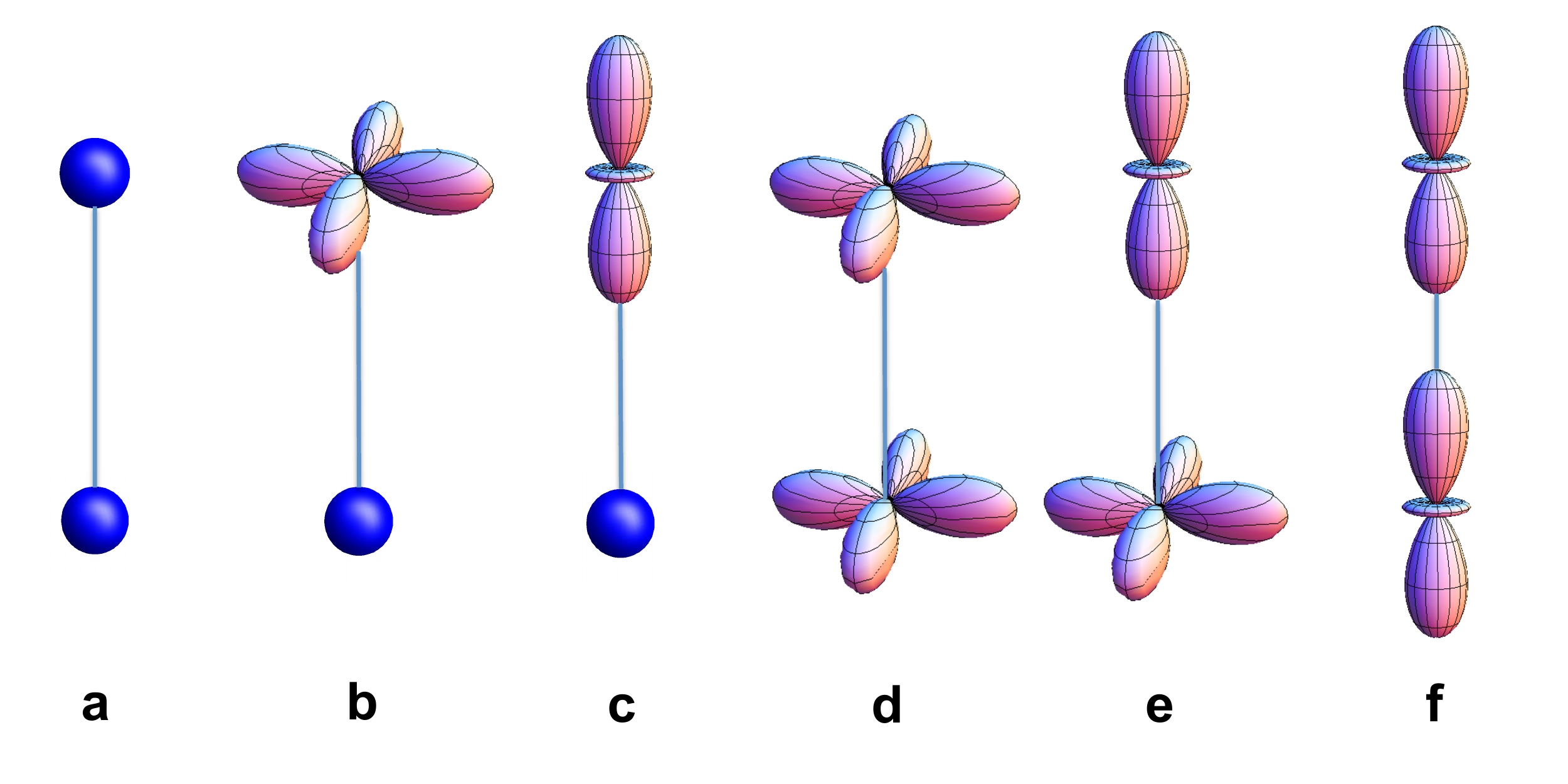}
\caption{{\bf Supplementary Information: A schematic showing the six inequivalent relative configurations for a pair of Co$^{3+}$ ions}. Panel {\bf a} shows a pair of nonmagnetic $S=0$ ions. {\bf b} and {\bf c} show a pair of $S=0$ and $S=1$ ions with $d_{x^2-y^2}$ and $d_{3z^2-r^2}$ orbitals, respectively. {\bf d}, {\bf e}, and {\bf f} show a pair of $S^z=1$ ions with orbital configurations $(d_{x^2-y^2}, d_{x^2-y^2})$, $(d_{3z^2-r^2}, d_{x^2-y^2})$, and $(d_{3z^2-r^2}, d_{3z^2-r^2})$, respectively.}
\label{configs}
\end{figure*}

Here we compute the exchange energies for a pair of Co$^{3+}$ ions on a nearest neighbor bond. Figure~\ref{configs} shows six inequivalent configurations of Co ions on a nearest-neighbor bond along the $z$-axis. The directional dependence of $e_g$ orbitals results in a highly anisotropic electron hopping, which in turn leads to an anisotropic exchange interaction. To simplify the calculation, we consider only the dominant $pd\sigma$ hopping between $e_g$ and $p$ orbitals of the O$^{2-}$ ions. The resultant exchange energy of a $z$-bond thus depends on whether the $d_{3z^2-r^2}$ orbitals are occupied. We expect $E_a = E_b = E_d = \epsilon_0$, $E_c = E_e = \epsilon_1$ and $E_f = \epsilon_2$. Here the subscripts of $\epsilon$ indicate the number of occupied $d_{3z^2-r^2}$ orbitals on the bond. For simplicity, we also neglect Hund's coupling.

We employ standard degenerate perturbation theory to calculate the bond energies up to fourth order in $t_{pd}$. For configurations {\bf a}, {\bf b} and {\bf d} in Fig.~\ref{configs}, the second and fourth order corrections to the bond energy are
\begin{eqnarray}
 \epsilon^{(2)}_0  = -  \frac{4 t_{pd}^2}{(\Delta_{pd}-U_{pp})},
\end{eqnarray}
and
\begin{eqnarray}
 \epsilon^{(4)}_0  &=& \frac{8 t_{pd}^4}{(\Delta_{pd}-U_{pp})^2} \Bigl[ \frac{2 }{(\Delta_{pd}-U_{pp})}  \\
 &&-  \frac{1}{(2 \Delta_{pd} -U_{pp})} -
 \frac{1}{(2 \Delta_{pd} + U_{dd}-U_{pp})} \Bigr]. \nonumber
\end{eqnarray}
Here, $U_{pp}$  and $U_{dd}$ are the Coulomb repulsion energies for double occupation of the $p_z$  and $d_{3 z^2 - r^2}$ orbitals respectively, while $\Delta_{pd}$ is the diagonal energy difference between the two orbitals. We have neglected corrections of order $J_{H}/U$ where $J_H$ is the Hund's exchange coupling. These perturbative corrections to the energy can be understood
to result from the virtual hopping of the $p$ electrons via the O ion to the neighboring Co sites and back. Similarly, we obtain the energy corrections
\begin{eqnarray}
 \epsilon^{(2)}_1  = - \frac{3t_{pd}^2}{(\Delta_{pd}-U_{pp})},
\end{eqnarray}
\begin{eqnarray}
 \epsilon^{(4)}_1  &=&- \frac{4t_{pd}^4}{(\Delta_{pd}-U_{pp})^2(2\Delta_{pd}+U_{dd}-U_{pp})} \nonumber \\
 && - \frac{4t_{pd}^4}{(\Delta_{pd}-U_{pp})^2(2\Delta_{pd}-U_{pp})} \\
 && - \frac{t_{pd}^4}{(\Delta_{pd}-U_{pp})^2U_{dd}} + \frac{9t^4}{(\Delta_{pd}-U_{pp})^3}
 \nonumber
\end{eqnarray}

\begin{eqnarray}
 \epsilon^{(2)}_2  &=& -  \frac{2 t_{pd}^2}{(\Delta_{pd}-U_{pp})}, \\
 \epsilon^{(4)}_2  &=&  \frac{4t_{pd}^4}{(\Delta_{pd}-U_{pp})^3}.
\end{eqnarray}

\subsubsection{Jahn-Teller interaction}
We compute the energy gain of a bond via Jahn-Teller coupling to the O$_6$ octahedra.
For simplicity, we adopt the so-called Einstein phonon model for the displacement $\mathbf u$ of the O ion, namely $E_{e} = k|\mathbf u|^2/2$. For a bond oriented along the $z$-axis, the JT couplings for the $d_{3z^2-r^2}$ and $d_{x^2-y^2}$ orbitals are $\gamma u_z$ and $-\gamma u_z/2$, respectively~\cite{khomskii03}, where $\gamma$ is a coupling constant. By integrating out the displacement $\mathbf u$, the energy gain for the six different configurations are
\begin{eqnarray}
	E_b = -\frac{\gamma^2}{8k}, \quad E_c = -\frac{\gamma^2}{2k}, \quad
	E_e = -\frac{9\gamma^2}{8k}, \nonumber \\ \nonumber \\ E_a = E_d = E_f = 0.
\end{eqnarray}

\subsection{Calculation of interaction potentials}

Here we derive the effective interaction potentials between a pair of  Co$^{3+}$ ions.
In particular we will focus on the exchange contributions to the potential energy. We introduce a pseudospin 1/2, $\bm\tau$, to describe the orbital degrees of freedom of the $S^z=+1$ ions such that $|\tau^z = \pm 1/2\rangle$ correspond to occupied orbital $|3z^2-r^2\rangle$ and $|x^2 - y^2\rangle$, respectively. As can be seen from the previous discussion, the exchange energy of a given bond depends  only on the occupation of the active orbital (e.g. $d_{3z^2-r^2}$ on a $z$-bond). We thus introduce an operator $P^{\nu}$ for the three different nearest neighbor bonds ($\nu=x$, $y$, or $z$) which projects onto the relevant orbital state. On introducing the following unit vectors
\begin{eqnarray}
	\hat\mathbf n_{x/y} = \pm\frac{\sqrt{3}}{2}\hat\mathbf x - \frac{1}{2}\hat\mathbf z,
	\quad\quad \hat\mathbf n_z = \hat\mathbf z,
\end{eqnarray}
the bond projection operators become $P^{\nu} = \frac{1}{2} + \bm\tau\cdot\hat\mathbf n_\nu$.
The exchange Hamiltonian is then given by
\begin{widetext}
\begin{eqnarray}
	H_{\rm ex} &=& \sum_{\nu}\sum_{\langle ij \rangle \parallel \nu} \Bigl\{ n_i n_j
	\bigl[ K_1\,(\bm\tau_i\cdot\hat\mathbf n_\nu)(\bm\tau_j\cdot\hat\mathbf n_\nu)
	+ K_2\,(\bm\tau_i+\bm\tau_j)\cdot\hat\mathbf n_\nu
	+ K_3\bigr] + \epsilon_0 (1-n_i)(1-n_j) \nonumber \\
	&& \quad + G_1\left[n_i (1-n_j) + (1-n_i) n_j\right]
	+ G_2 \left[ n_i(1-n_j)\, \bm\tau_i\cdot\hat\mathbf n_\nu + (1-n_i)n_j \,\bm\tau_j\cdot\hat\mathbf n_\nu\right]
	\Bigr\}.
\end{eqnarray}
\end{widetext}
Here $K_1 = \epsilon_2 + \epsilon_0 - 2\epsilon_1$, $K_2 = (\epsilon_2 - \epsilon_0)/2$,
$K_3 = (\epsilon_2 + \epsilon_0 + 2\epsilon_1)/4$, $G_1 = (\epsilon_1 + \epsilon_0)/2$, and $G_2 = (\epsilon_1 - \epsilon_0)$.

In the regime dominated by nonlinear JT interactions, quantum fluctuations of the orbitals are quenched by the lattice distortion. We thus replace the pseudospin operator by a classical O(2) vector of length $|\bm\tau|=1/2$. The local elongation of the O$_6$ octahedron aligns the $\bm\tau_i$ vector to one of the three directions $\hat\mathbf n_\nu$. Each corresponds to the three Potts states discussed in the main text. Noting that $\bm\tau_i\cdot\hat\mathbf n_{\nu} = 1/2$ for $s_i = \nu$ and $-1/4$ otherwise, the effective nearest neighbor interaction potentials in Equation~(\ref{eq:H2}) can be easily read off
from $H_{\rm ex}$ as
\begin{eqnarray}
	V_1 = \frac{\mathcal{J}}{16}, \quad V^a_1 = \frac{3\mathcal{J}}{16}, \quad V^f_1 = \frac{15\mathcal{J}}{16},
\end{eqnarray}
where the exchange constant $\mathcal{J}$ is given by
\begin{eqnarray}
	\mathcal{J} &=& \epsilon_2 + \epsilon_0 - 2 \epsilon_1 \\
	&=& \frac{2t_{pd}^4}{(\Delta_{pd}-U_{pp})^2}\left[\frac{1}{\Delta-U_{pp}} + \frac{1}{U_{dd}}\right]. \nonumber
\end{eqnarray}
and the correction to the gap energy is
\begin{eqnarray}
	\Delta_{\rm ex} = 3(\epsilon_1 - \epsilon_0) = \frac{3 t_{pd}^2}{(\Delta_{pd}-U_{pp})} + \mathcal{O}(t_{pd}^4).
\end{eqnarray}

\end{document}